\documentclass[a4paper,12pt]{article}
\usepackage{epsfig}

\def\lsim{\raise0.3ex\hbox{$\;<$\kern-0.75em\raise-1.1ex\hbox{$\sim\;$}}}
\def\gsim{\raise0.3ex\hbox{$\;>$\kern-0.75em\raise-1.1ex\hbox{$\sim\;$}}}

\def\bmat{\left(\begin{array}}
\def\emat{\end{array}\right)}
\def    \be            {\begin{equation}}
\def    \ee            {\end{equation}}
\def    \bea           {\begin{eqnarray}}
\def    \eea           {\end{eqnarray}}
\def    \nn            {\nonumber}

\begin{document}

\title{Supersymmetric dark matter and neutralino-nucleon cross
section
}

%
%
%
%
%

\rightline{FTUAM 01/10}
\rightline{IFT-UAM/CSIC-01-16}
\rightline{SUSX-TH/01-021}
\rightline{hep-ph/0105180}
\rightline{May 2001}   

\renewcommand{\thefootnote}{\fnsymbol{footnote}}
\setcounter{footnote}{0}
\vspace{0.0cm}
\begin{center}
{\large{\bf Supersymmetric dark matter and neutralino-nucleon 
\\ cross section \footnote{Extension of a talk given at Cairo International
Conference on High Energy Physics (CICHEP 2001), January 9-14.}\\[5mm]}}

\sc{D.G. Cerde\~no}\\
{\small
{\it Departamento de F\'{\i}sica
Te\'orica C-XI, Universidad Aut\'onoma de Madrid,\\
Cantoblanco, 28049 Madrid, Spain. \\[5mm]}}
\sc{S. Khalil}\\
{\small
{\it Centre for Theoretical Physics, University of Sussex, 
Brighton BN1 9QJ, U.K.\\
\it  Ain Shams University, Faculty of Science, Cairo
11566, Egypt.}}\\[5mm]
\sc{C. Mu\~noz}\\
{\small
{\it  Departamento de F\'{\i}sica
Te\'orica C-XI and Instituto de F\'{\i}sica Te\'orica C-XVI,\\ 
 Universidad Aut\'onoma de Madrid,
Cantoblanco, 28049 Madrid, Spain. \\}}

\end{center}



\renewcommand{\thefootnote}{\alph{footnote}}
\setcounter{footnote}{0}

{\small We review the direct detection of supersymmetric dark
matter in the light of recent experimental results. In particular, 
we show that regions in the parameter space of
several supergravity scenarios with a 
neutralino-nucleon cross section of the order of $10^{-6}$ pb,
i.e., where current dark matter detectors are sensitive, can be
obtained. 
These are scenarios with large 
$\tan \beta$, with non-universal soft supersymmetry-breaking terms, 
with multi-TeV masses for scalar superpartners known as 
`focus point' supersymmetry, 
and finally scenarios with intermediate unification 
scale which appear naturally in some superstring constructions.}

\section{\large Introduction}
One of the strongest motivations for physics beyond the standard model is the
existence of dark matter. Substantial evidences exist suggesting that most
of the mass in the Universe is some non-luminous matter, the so called
`dark matter',
of as yet unknown composition. Currently the most convincing observational 
evidence for the existence of dark matter comes from the analysis
of rotation curves of spiral Galaxies, i.e., measurements of the velocity 
of isolated stars or gas clouds which orbit in the outer parts
of spiral Galaxies. It
has been noted that there is not enough luminous matter in those
Galaxies to account for their observed rotation curves \cite{rotation}. 

The detailed analysis of rotation curves offers convincing evidence that 
90\% or more of the mass of the Galaxies is dark. 
Other evidences arise from large scale measurements, as e.g. the
motions
of cluster members Galaxies.
A key question is then, what 
could this dark matter be? We don't know the answer
yet, but we do know that not all of it can be ordinary (baryonic) matter,
since measured abundances of helium, deuterium and lithium in the 
scenario of big-bang nucleosynthesis 
impose a strong upper bound on the baryon density in the Universe. This 
density
is too small to account for the whole dark matter in the Universe. 
The conclusion is that baryonic objects, such as e.g. MACHOs, 
can be components of the dark matter, but more candidates are 
needed \cite{Sadoulet}.

Fortunately, 
particle physics offers various candidates for (non-baryonic) dark matter,
all of which would indicate new physics beyond the well tested standard model
of particle physics \cite{Drees}. For example, 
long-lived or stable weakly-interacting massive 
particles (WIMPs) can remain from the earliest moments of the Universe in 
sufficient number to account for a significant fraction of relic density. 
These particles would form not only a background density in the Universe, but
also would cluster gravitationally with ordinary stars in the galactic halos.

This raises the hope of detecting relic 
WIMPs directly, by observing their elastic scattering on  
target nuclei through nuclear recoils.
Since WIMPs interact with ordinary matter with very roughly weak
strength, and assuming that their masses are of the order of  
weak scale (i.e., between 10 GeV and a few TeV),
it is natural to expect a WIMP-nucleus cross section of the same order
as that of a weak process, which is around 1 pb.
This would imply a WIMP-nucleon cross section around   
$10^{-8}$ pb, too low to be detected by current dark matter
experiments,
DAMA, CDMS and UKCDM,
which are sensitive to a cross section around $10^{-6}$ pb.
Surprisingly, 
the DAMA collaboration reported recently \cite{experimento1}
data favouring the existence of a 
WIMP signal in their search for annual modulation.
When 
uncertainties as e.g. the WIMP velocity or possible bulk halo rotation,
are included, it was claimed that the preferred range of parameters
is  (at 4$\sigma$ C.L.) $10^{-6}$ pb $\lsim \sigma\lsim 10^{-5}$ pb 
for a WIMP mass 30 GeV $\lsim m \lsim 200$ GeV.
Unlike this spectacular result, the CDMS collaboration 
claims to have excluded \cite{experimento2} regions of the DAMA 
parameter space.

In any case, due to these and other projected experiments, 
it seems very plausible that the dark matter 
will be found in the near future. 
In this situation, and assuming that the dark matter 
is a WIMP, it is natural to wonder how big 
the cross section for its direct detection can be.
The answer to this 
question depends on the particular WIMP considered.
The leading candidate in this class is the lightest 
neutralino \cite{kami}, a particle 
predicted by the supersymmetric (SUSY) extension of the standard model.
In this paper we 
critically reappraise the known SUSY scenarios based on 
neutralinos as dark-matter candidates, and in particular
the scenarios constructed recently in order to enhance the neutralino-nucleon
cross section. 
This is the case of scenarios with large 
$\tan \beta$ \cite{Bottino}-\cite{Mario}, with
non-universal soft SUSY-breaking terms \cite{Bottino,Arnowitt,Nath2}, 
with multi-TeV masses
for scalar superpartners known as `focus point' 
supersymmetry \cite{focus}, and
finally scenarios with intermediate unification 
scale \cite{muas}-\cite{nosotros}.


\section{\large Supersymmetric predictions for the neutralino-nucleon cross section}

In SUSY models, $R$-parity is often imposed to avoid weak scale 
proton decay or lepton number violation. Imposing this symmetry yields 
remarkable phenomenological implications. SUSY particles are produced 
or destroyed only in pairs and, as a consequence, the lightest 
supersymmetric particle (LSP) 
is absolutely stable.
The former implies that
a major signature 
for $R$-parity conserving models is represented by events with missing 
energy (for instance, $e^+ e^- \rightarrow \mathrm{jet}+ \mathrm{ missing\ 
energy}$). The latter implies that the LSP 
might constitute a possible candidate for dark matter. 
Concerning this point, it is remarkable that in most of the parameter
space of SUSY models the LSP is an electrically neutral 
(also with no strong interactions) particle, called neutralino.
This is welcome since otherwise the LSP 
would bind to nuclei and would be excluded as a candidate
for dark matter from unsuccessful searches for exotic heavy 
isotopes \cite{isotopes}.

In the simplest SUSY extension of the standard model, the so-called
minimal supersymmetric standard model (MSSM)
there are four neutralinos, $\tilde{\chi}^0_i~(i=1,2,3,4)$, since they
are the physical 
superpositions of the fermionic partners of the neutral electroweak 
gauge bosons, 
called bino ($\tilde{B}^0$) and wino ($\tilde{W}_3^0$), and of the
fermionic partners of the  
neutral Higgs bosons, called Higgsinos ($\tilde{H}^0_u$, 
$\tilde{H}_d^0$). 
Therefore the lightest neutralino, $\tilde{\chi}^0_1$, will be the 
dark matter candidate.
The neutralino mass matrix 
with the conventions for gaugino and Higgsino masses in the Lagrangian,
${\cal L} =\frac{1}{2}\sum_{a} M_a {\lambda}_a \lambda_a + 
\mu \tilde{H}^0_u \tilde{H}^0_d$ + h.c.,
is given by
\begin{equation}
{\small \bmat{cccc}M_1 & 0 &-M_Z\cos\beta \sin\theta_W &M_Z\sin\beta 
\sin\theta_W \\ 0 & M_2 & M_Z \cos \beta \cos\theta_W & -M_Z \sin \beta 
\cos\theta_W \\ -M_Z \cos \beta \sin\theta_W & M_Z \cos \beta\cos\theta_W&
0& -\mu \\ M_Z \sin \beta \sin\theta_W & -M_Z \sin \beta\cos\theta_W&-\mu 
& 0 \emat,}
\label{mass}
\end{equation}
in the above basis 
($\tilde B^0=-i{\lambda}'$, $\tilde W_3^0=-i{\lambda}^3$, $\tilde H_u^0$, 
$\tilde H_d^0$). 
Here $M_1$ and $M_2$ are the soft bino and wino masses respectively, 
$\mu$ is the Higgsino mass parameter 
and 
$ \tan\beta= \langle H_u^0\rangle/\langle H_d^0\rangle$ 
is the ratio of Higgs vacuum expectation values. 
We parameterize the gaugino and Higgsino content
of the lightest neutralino according to 
\begin{equation}
\tilde{\chi}^0_1 = N_{11} \tilde{B}^0 +N_{12} \tilde{W}_3^0 +
N_{13} \tilde{H}^0_d + N_{14} \tilde{H}^0_u\ .
\label{lneu}
\end{equation}
It is commonly defined that $\tilde{\chi}^0_1$ is mostly gaugino-like 
if $P\equiv \vert N_{11}\vert^2 + \vert N_{12} \vert^2 > 0.9$, Higgsino-like
if $P<0.1$, and mixed otherwise.

\begin{figure}[t]
\begin{center}
\epsfig{file= 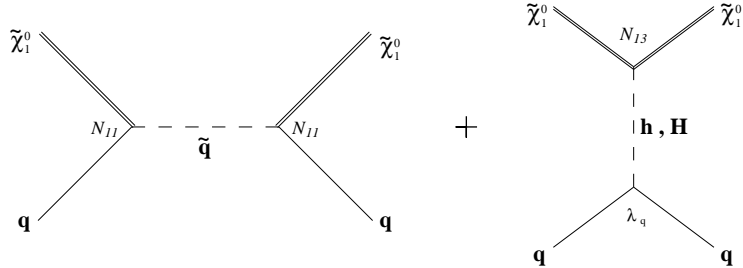,
height=3.5cm,angle=0}
\end{center}
\caption{Feynman diagrams contributing to neutralino-nucleon cross section.}
\label{Feynman}
\end{figure}

The relevant effective Lagrangian describing
the elastic $\tilde{\chi}^0_1$-nucleon scattering in the MSSM is given by
\begin{equation}
\mathcal{L}_{eff} = \sum_{q} \left(\alpha_{q} \bar{\chi}\chi \bar{q} q
+
\beta_q \bar{\chi} \gamma^{\mu}  \gamma^{5} 
\chi \bar{q} \gamma_{\mu}  \gamma^{5} q\right)\ ,
\end{equation}
where  $\alpha_q$, $\beta_q$ are given e.g. in ref.\cite{Ellis} 
with the sign conventions for Yukawa couplings in the Lagrangian,
${\cal L}=-h_u H_u^0 \bar u_L u_R -h_d H_d^0 \bar d_L d_R 
- h_e H_d^0 \bar e_L e_R $ + h.c.,
and the sum 
runs over the six quarks. 
The contribution of the scalar (spin-independent) interaction,
the one proportional to $\alpha_q$, to the $\tilde{\chi}^0_1$-nucleon 
cross section is generically larger than the spin-dependent
interaction,
the one proportional to $\beta_q$.
One can concentrate then on the scalar 
$\tilde{\chi}^0_1$-nucleon
cross section. This is given by
\begin{equation}
\sigma
= \frac{4 m_{r}^2}{\pi} f^2\ ,
\label{crosssection}
\end{equation}
where $m_{r}$ is the reduced $\tilde{\chi}^0_1$ mass and
\begin{equation}
\frac{f}{m} = \sum_{q} f_q \frac{\alpha_q}{m_q}
\end{equation}
with $m$ the mass of the nucleon and $m_q$ the mass of the quarks. 
The scalar 
coefficients $\alpha_{q}$ include contributions from 
squark ($\tilde q$) exchange and neutral Higgs 
$(h, H)$ exchange, as illustrated in Fig.~\ref{Feynman}.  
The numerical values of the hadronic matrix elements $f_q$ 
for the proton are as follows \cite{Ellis}:
\begin{equation}
f_u= 0.020 \pm 0.004\ , \hspace{0.5cm} f_d= 0.026 \pm
0.005\ ,
\hspace{0.5cm} f_s= 0.118 \pm 0.062\ ,
\label{central}
\end{equation}
and \footnote{Larger values,
as for example $f_s=0.455$, have also been used in the 
literature, see e.g. ref.~\cite{Arnowitt}.} 
$f_{c,b,t}= \frac{2}{27} (1-\sum_{q=u,d,s} f_q)$.
For the neutron the value of $f_s$ is the same and therefore
the scalar cross sections for both, protons and neutrons, are
basically equal. 

The cross section for the elastic scattering of relic neutralinos on 
protons and neutrons has been examined exhaustively in the 
literature \cite{kami}.   
This is for example the case in
the framework of minimal supergravity (mSUGRA).
Let us recall that in this framework one makes several 
assumptions. In particular, 
the scalar mass parameters, the gaugino mass parameters, and the 
trilinear couplings, which are generated once SUSY is broken through
gravitational interactions, are universal at the 
grand unification 
scale, $M_{GUT} \approx 2\times 10^{16}$ GeV.
They are denoted by $m_0$, $M_{1/2}$, and $A_0$ respectively. 
Likewise, 
radiative electroweak symmetry breaking is imposed, i.e., $\vert \mu 
\vert$ is determined by the minimization of the Higgs effective 
potential. This implies 
\begin{equation}
\mu^2 = \frac{m_{H_d}^2 - m_{H_u}^2 \tan^2 \beta}{\tan^2 \beta -1 } - 
\frac{1}{2} M_Z^2\ .
\label{electroweak}
\end{equation} 
With these assumptions, the mSUGRA framework  allows four free 
parameters: $m_0$, $M_{1/2}$, $A_0$, and $\tan \beta$. In addition, the
sign of $\mu$ 
remains also undetermined.  

\begin{figure}[t]
\begin{center}
\epsfig{file= 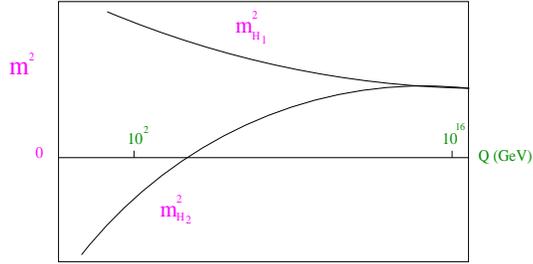,width=7cm,height=3.5cm,angle=0}
\end{center}
\caption{
Running of the soft Higgs masses-squared with energy.}
\label{run}
\end{figure}

It was observed (for a recent re-evaluation see ref.\cite{Ellis})
that for low and moderate values of
$\tan \beta$ the predicted scalar neutralino-proton cross sections
are well below the accessible experimental regions.
To understand this result qualitatively, we show schematically in
Fig.~\ref{run} 
the well known evolution of $m_{H_d}^2$ and $m_{H_u}^2$ with the
scale\footnote{Let us remark that, given the convention used throughout
this review for gaugino masses in the Lagrangian, 
${\cal L} =\frac{1}{2}\sum_{a} M_a {\lambda}_a \lambda_a$ + h.c., we
are using the renormalization group equations (RGEs) obtained e.g.
in ref.\cite{rges}, but with an opposite sign in the
gaugino contributions to the RGE's of the $A_0$ parameters.}. 
Since $m_{H_u}^2$ becomes large and negative (notice also
that $m_{H_d}^2$, neglecting
bottom and tau Yukawa couplings, becomes positive), $|\mu|$ due
to relation (\ref{electroweak}) 
becomes also large, in particular much larger than
$M_1$ and $M_2$. Thus, as can be easily understood from 
eqs.(\ref{mass}) and (\ref{lneu}), 
the lightest neutralino will be mainly gaugino, and in particular
bino since at low energy $M_1\approx \frac{1}{2}M_2$.

\begin{figure}[t]
\begin{center}
\epsfig{file= 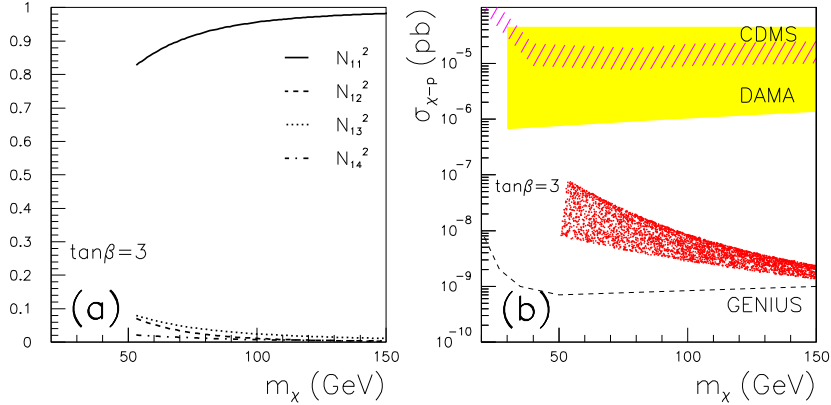,
height=7cm
}
\end{center}

\vspace{-0.5cm}

\caption{(a) Gaugino-Higgsino components-squared
of the lightest neutralino as a function of its mass.
(b) 
Scatter plot of the neutralino-proton cross section
as a function of 
the neutralino mass for the scenario discussed in the text.
DAMA and CDMS current 
experimental limits and projected GENIUS limits are shown.
}
\label{ellisfig}
\end{figure}

We show this fact in Fig.~\ref{ellisfig}a, 
where for $\tan\beta=3$ the gaugino-Higgsino 
components-squared $N_{1i}^2$ of the lightest neutralino as a function
of its mass $m_{\tilde{\chi}_1^0}$ 
are exhibited. Here we are using as an example $m_0=150$ GeV, and
$M_{1/2}$ is essentially fixed for a given
$m_{\tilde{\chi}_1^0}$.
Note that 
$N_{11}$ is 
extremely large and therefore $P\gsim 0.9$.
Then, the scattering channels through Higgs exchange shown in 
Fig.~\ref{Feynman} are suppressed (recall that the Higgs-neutralino-neutralino
couplings are proportional to $N_{13}$ and $N_{14}$) and therefore
the cross section is small.
As a matter of fact, the scattering channels through squark 
exchange, shown also in Fig.~\ref{Feynman}, 
are also suppressed by the mass of the squarks. Indeed
in this limit the cross section (\ref{crosssection}) can be approximated as
\bea
\sigma_{\tilde\chi_1^0-p}
\approx {m_r^2}\ \frac{\alpha^2}{\left(m_{\tilde s}^2-
m_{\tilde{\chi}_1^0}^2\right)^2}\  f_s^2\   |N_{11}|^4
\ .
\label{cross}
\eea
Thus it can be roughly estimated to be 
$\sigma_{\tilde{\chi}_1^0-p}\approx 10^{-8}$ pb,
for $\alpha\approx 10^{-2}$, 
$f_s\approx 10^{-1}$,
$m_r\approx 1$ GeV and 
$m_{\tilde s}\approx 300$ GeV. This is precisely the value of
the cross section 
expected for a weak process as discussed in the Introduction.

In Fig.~\ref{ellisfig}b we show the cross section
$\sigma_{\tilde{\chi}_1^0-p}$ as a function of the neutralino mass
$m_{\tilde{\chi}_1^0}$ for
$\tan \beta=3$.
We also choose to plot only the case $\mu > 0$ since for negative values of
$\mu$ the cross sections are much
smaller\footnote{It is also worth
noticing that constraints
coming from
the $b\rightarrow s\gamma$ process highly reduce
the $\mu<0$ parameter space. In addition,
recent measurements of the anomalous magnetic moment of the
muon \cite{marci} seems to exclude all cases with $\mu<0$.}.
Fig.~\ref{ellisfig}b has been obtained
using formula (\ref{crosssection}) with the central
values for the hadronic matrix elements given by
eq.(\ref{central}). We will use these values throughout this
review\footnote{It is worth mentioning here that values for the hadronic matrix
elements as those discussed in footnote {\it a}
will increase in one order of magnitude the results of this and
the other figures in this review,
since their contribution appears raised to the square
in eq.(\ref{crosssection}).}.
Likewise we impose
the present
bounds coming from accelerators.
These are
LEP and Tevatron bounds on SUSY masses and
CLEO $b\to s\gamma$ branching ratio measurements.
For the Higgs mass we use the present experimental
lower limit for $\tan\beta=3$ \cite{lep}, i.e. $m_h>95$ GeV.
Although, for the values of $\tan\beta$ that we will
consider in the next section, $\tan\beta=10$,
this limit is considerably weaker \cite{lep},
$m_h>80$ GeV, we will still use $m_h>95$ GeV.
Concerning the parameter space of the figure, we have used
30 GeV $\leq m_0 \leq 550$ GeV, where the lower bound 
is in order to avoid the stau being the LSP for all values of 
$m_{\tilde{\chi}_1^0}$.
Note that the curve associated to $m_0=550$ GeV 
corresponds to the minimum values of 
$\sigma_{\tilde{\chi}_{1}^{0}-p}$ in the figure.
On the other hand, since the cross sections are not very sensitive to 
the specific values of $A_0$ in a wide 
range (we have checked that 
this is so for $\mid A_0/M_{1/2}\mid \lsim 1$), we fix
$A_0 = M_{1/2}$ in the figure.
As mentioned above 
the gaugino mass
$M_{1/2}$ is essentially fixed for a given $\tan\beta$ and 
$m_{\tilde{\chi}_1^0}$. 
For this figure we have to take
140 GeV $\lsim M_{1/2} \lsim 350$ GeV,
where the lower bound, corresponding to $m_{\tilde{\chi}_1^0}\geq 55$ GeV,
is due to the experimental bound
on the lightest chargino mass
$m_{\tilde\chi_1^{\pm}}>90$ GeV. 
Negative values of $M_{1/2}$, not shown in the
figure,
correspond to smaller cross sections.



As we can see in the figure
$\sigma_{\tilde{\chi}_1^0-p} \lsim 10^{-7}$ pb, and
therefore
for these values of the parameters
we would have to wait in principle for the 
projected GENIUS detector \cite{GENIUS} to be able to
test the neutralino as a dark-matter candidate.

\section{\large Large neutralino-nucleon cross section in supergravity scenarios}
 
Recently there has been some theoretical 
activity \cite{Bottino}-\cite{nosotros} trying to obtain regions
in the parameter space of supergravity (SUGRA) scenarios 
compatible with the sensitivity of
current dark matter detectors, DAMA and CDMS. 
The key point in most of these scenarios 
to carry it out consists of reducing the value of $|\mu|$.
Following the discussion below eq.(\ref{electroweak}) in the previous section, 
the smaller $|\mu|$ is,
the larger the Higgsino components of the lightest neutralino become.
Eventually, $|\mu|$ will be of the order of $M_1$, $M_2$
and $\tilde{\chi}_1^0$ will be a mixed Higgsino-gaugino state.
Indeed scattering channels through Higgs exchange in 
Fig.~\ref{Feynman}
are now important and their contributions to the cross section
(\ref{crosssection}) can
be schematically approximated as 
\bea
\sigma_{\tilde\chi_1^0-p}
\approx {m_r^2}\
\frac{\lambda_s^2 \alpha}{m_h^4}\
f_s^2\   |N_{13} N_{11}|^2 
\ ,
\label{cross2}
\eea
where $\lambda_s$ is the strange quark Yukawa coupling and
$m_h$ represent the Higgs masses.
With the same rough estimate as in eq.(\ref{cross}),
one obtains $\sigma_{\tilde{\chi}_1^0-p}\approx 10^{-6}$ pb for 
$m_h\approx 100$ GeV.

In this section we review the different SUGRA scenarios 
that can be found in the
literature in order to enhance the neutralino-proton cross section
$\sigma_{\tilde\chi_1^0-p}$ to be of the order of $10^{-6}$ pb.

\subsection{\normalsize Scenario with large $\tan \beta$}

\begin{figure}[t]
\begin{center}
\epsfig{figure=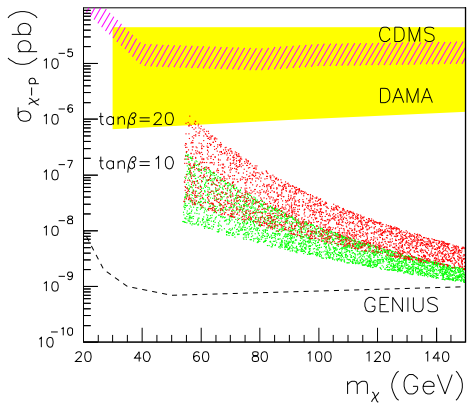, 
height=7cm}
\end{center}

\vspace{-0.5cm}

\caption{The same as in Fig.~\ref{ellisfig}b but for 
$\tan\beta=10$ and 20.
}
\label{ellisfig2}
\end{figure}

In the previous section we showed in Fig.~\ref{ellisfig}b
the neutralino-proton cross section for $\tan\beta=3$, in the framework 
of mSUGRA.
Here we show the same in Fig.~\ref{ellisfig2} but for larger values of
$\tan\beta$,
in particular for  $\tan\beta=10$ (green points) and 20 (red points).
We see 
that the cross section increases when the value
of $\tan\beta$ increases.
For moderate values of $\tan\beta$, 
the reason being that the top(bottom) Yukawa coupling
which appears in the RGE
for $m_{H_u}^2$($m_{H_d}^2$)
decreases(increases) since it is proportional
to $\frac{1}{\sin\beta}$($\frac{1}{\cos\beta}$).
This implies that the negative(positive) contribution 
$m_{H_u}^2$($m_{H_d}^2$) to $\mu^2$ in eq.(\ref{electroweak}) 
is less important, and 
therefore $|\mu|$ decreases. 

However, for $\tan\beta \gsim 10$, the value of $\mu^2$ is very
stable with respect to variations of $\tan\beta$.
This is due to the fact 
that $\mu^2\approx -m_{H_2}^2 - \frac{1}{2}M_Z^2$ (see
eq.(\ref{electroweak})). Since $\sin\beta\approx 1$, the top Yukawa
coupling is stable and therefore the same conclusion is obtained for
$m_{H_2}^2$ and $\mu^2$. Thus, 
the reason for the cross section to increase
when $\tan\beta$ increases
cannot be now the increment of the Higgsino components of the LSP.
Nevertheless there is a second effect in the cross section 
which is now the dominant one: the contribution of the down-type
quark Yukawa couplings (see eq.(\ref{cross2})) 
which are proportional to $\frac{1}{\cos\beta}$.

It was in fact
pointed out in refs.\cite{Bottino,Arnowitt} that the large
$\tan\beta$ regime allows regions where 
$\sigma_{\tilde{\chi}_1^0-p}\approx 10^{-6}$ pb is reached. 
We can see in Fig.~\ref{ellisfig2} that this is so for 
$\tan\beta \gsim 20$.

Very large values of 
$\tan\beta$, 
like $\tan \beta \simeq 50$ which 
correspond approximately to the unification of the tau and top Yukawa 
couplings at $M_{GUT}$,
have also been considered \cite{Mario}.
Although it was found, as expected, that the 
cross section is enhanced,
the 
well known
experimental limits coming from 
$b \to s \gamma$ for large $\tan\beta$, 
lead to severe constraints on the parameter space. In particular,
these constraints imply $\sigma_{\tilde{\chi}_1^0-p} \lsim 
10^{-6}$ pb. 


\subsection{\normalsize Focus point supersymmetry scenario}

Another possibility 
to obtain a large 
neutralino-nucleon cross section 
arises in the so-called focus point 
supersymmetry scenario.
This has been proposed \cite{Feng} 
in order to avoid dangerous SUSY contributions to 
flavour and CP violating effects.
The idea is to assume the existence of squark and sleptons with
masses which can be taken well 
above\footnote{Of course this scenario rules out SUSY
as an explanation of the
possible deviation in the muon anomalous magnetic moment
from the standard model prediction \cite{marci}.} 1 TeV.
It has also been argued that this situation produces no loss of naturalness.

The implications of focus point supersymmetry for neutralino dark matter
have been considered in ref.\cite{focus}.
In particular, it was pointed out that for 
$m_0 > 1$ TeV, unlike the usual cases with $m_0 < 1$ TeV,  
the lightest neutralino is a gaugino-Higgsino 
mixture over much of parameter space.
Let us recall from the previous subsection 
that for moderate and large values of $\tan\beta$,
$\mu^2$ in eq.(\ref{electroweak}) can be approximated as 
%
\begin{equation}
\mu^2 \approx - m_{H_u}^2 - \frac{1}{2} M_Z^2. 
\label{approxmu}
\end{equation}
Thus 
for $m_0 > 1$ TeV, 
$m_{H_u}^2$ becomes less negative, and therefore $|\mu|$ decreases.
(For $\tan \beta < 5$, $\mu $ becomes  
sensitive to $m_{H_d}^2$, 
and as $m_0$ increases $|\mu|$ also increases, and so there is no mixed   
gaugino-Higgsino region for small $\tan \beta$.)
However, as discussed in ref.\cite{focus}, one still needs large values for 
$\tan \beta$ (of the order of 50) in order to have 
$\sigma_{\tilde{\chi}_1^0-p} \lsim 
10^{-6}$ pb. 

\subsection{\normalsize Scenario with non-universal soft terms}

The soft SUSY-breaking terms can have in general a
non-universal structure in the MSSM. Such structure can
be derived from SUGRA \cite{dilaton}. 
For the case of the observable scalar masses,
this is due to the non-universal couplings
in the K\"ahler potential
between the hidden sector fields breaking SUSY and the
observable sector fields.
For the case of the gaugino masses, this is due to the
non-universality of the gauge kinetic functions associated to the
different gauge groups.
Explicit string constructions, whose low-energy limit is SUGRA,
exhibit these properties \cite{dilaton}.
 
It was shown recently, in the context of SUGRA, 
that non-universality allows to increase \cite{Bottino,Arnowitt,Nath2}
the neutralino-proton cross section for moderate values of $\tan\beta$.
This can be carried out with non-universal 
scalar masses
and/or gaugino masses, 
as we will discuss below.\\

\hspace{-0.65cm}{\it $(i)$ Non-universal scalar masses}\\

\begin{figure}[t]
\begin{center}
\epsfig{figure=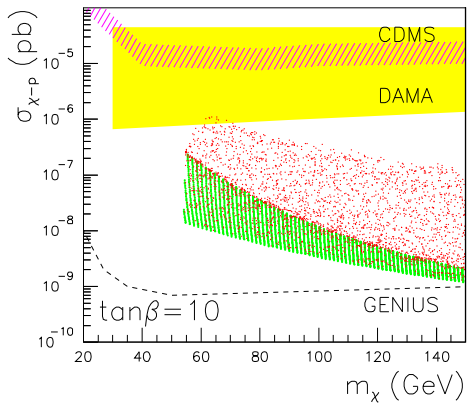, 
height=7cm}
\end{center}

\vspace{-0.5cm}

\caption{
The same as in Fig.~\ref{ellisfig}b
but for $\tan{\beta}=10$ and using non-universal soft 
scalar masses 
as explained in the text.
The universal case (green region) 
is shown for comparison.
}
\label{arnowit}
\end{figure}

Non-universality in the Higgs sector, concerning dark matter, was studied in 
refs.\cite{Ole,Fornengo,Bottino}. Subsequently, non-universality in
the sfermion sector was added in the analysis \cite{arna,Arnowitt}.
In order to avoid potential problems
with flavour changing neutral currents, it was 
assumed that
the first two generations of squarks
and sleptons have a common scalar mass
$m_0$ at $M_{GUT}$, and 
non-universalities were allowed in the third generation and
Higgs sector.
Thus the soft masses can be parameterized as follows:
\begin{eqnarray}
m_{H_{d}}^2&=&m_{0}^{2}(1+\delta_{1}), \quad m_{H_{u}}^{\ 2}=m_{0}^{2}
(1+ \delta_{2}),\nonumber \\
m_{Q_{L}}^2&=&m_{0}^{2}(1+\delta_{3}), \quad m_{u_{R}}^{\ 2}=m_{0}^{2}
(1+\delta_{4}), \nonumber\\
m_{e_{R}}^2&=&m_{0}^{2}(1+\delta_{5}),  \quad m_{d_{R}}^{\ 2}=m_{0}^{2}
(1+\delta_{6}), \nonumber\\
m_{L_{L}}^2&=&m_{0}^{2}(1+\delta_{7}),     
\end{eqnarray}
where 
$Q_{L}=(\tilde{t}_{L}, \tilde{b}_{L})$, $L_{L}=(\tilde{\nu}_{L},
\tilde{\tau}_{L})$, $u_{R}=\tilde{t}_{R}$, $e_{R}=\tilde{\tau}_{R}$, and 
it is assumed that $ -1 \leq \delta_{i} \leq 1 $.

As explained at the beginning of this section~3,
$\mu^{2}$ is one of the important parameters when
computing
the neutralino-nucleon cross section.
Its value is determined by condition (\ref{electroweak}) and can
be significantly reduced for some choices of $\delta$'s. We can have
a qualitative
understanding of the effects of the $\delta$'s on $\mu$ from
the following.
First, when $m_{H_u}^2$($m_{H_d}^2$) at $M_{GUT}$ increases(decreases)
its negative(positive) contribution at low energy in eq.(\ref{electroweak})
is less important. Second, when $m_{Q_{L}}^2$ and $m_{u_{R}}^2$
at $M_{GUT}$ decrease, due to their contribution proportional
to the top Yukawa coupling in the RGE of $m_{H_u}^2$ the negative
contribution of the latter to $\mu^2$ is again less important.
%
%
Thus one can deduce that 
$\mu^2$ will be reduced (and hence $\sigma_{\tilde{\chi}_1^0-p}$ increased)
by choosing $\delta_3, \delta_4, \delta_1 < 0$ and $\delta_2 >0$.

Following this analysis \cite{Bottino,Arnowitt} we show in Fig.~\ref{arnowit},
for $\tan\beta=10$,
a scatter plot of
$\sigma_{\tilde{\chi}_{1}^{0}-p}$ as a function of $m_{\tilde{\chi}_{1}^{0}}$
for a scanning of the parameters as follows:
$0 \leq \delta_2 \leq 1$, $-1 \leq \delta_{1,3,4} \leq 0$.
The other $\delta'$s are not so important in the computation and we take
them vanishing.
We are also taking $A=M_{1/2}$, 30 GeV $\leq m_0\leq 550 $ GeV as in
Fig.~\ref{ellisfig}b,
and we have to take now 140 GeV $\lsim M_{1/2} \lsim
450$ GeV.
For comparison we superimpose also the region (green area) 
obtained in Fig.~\ref{ellisfig2} for $\tan\beta=10$
with universality, $\delta_i=0$.
We see that non-universal scalar masses can help in increasing the values of
$\sigma_{\tilde{\chi}_{1}^{0}-p}$.
In fact non-universality in the Higgs sector gives the most important
effect, and including the one in the sfermion sector the cross
section only increases slightly.
Of course, as discussed in detail in Subsection 3.1, for
larger values of $\tan{\beta}$ one can get 
larger cross sections.\\

\hspace{-0.65cm}{\it $(ii)$ Non-universal gaugino masses}\\

The effects of the non-universality of gaugino masses on dark matter
in SUGRA scenarios have been studied 
in ref.\cite{Yamaguchi,Nath2,Nelson}.
In particular, in ref.\cite{Nath2}, the authors analyze in detail 
the neutralino-proton cross section in the case of
$SU(5)$ unified models with interesting results, 
finding that the cross sections 
can increase. 

Here we are interested in the case of the standard model gauge group
in order to compare our results with both, the universal case discussed
in Sections 2 and 3.1 and the case with non-universal scalar masses
discussed above.
Let us then parameterize the soft gaugino masses at $M_{GUT}$ as follows:
%
\begin{eqnarray}
M_{1}&=&M_{1/2}(1+\delta'_{1}), \quad M_2=M_{1/2}
(1+ \delta'_{2}), \quad M_3=M_{1/2}(1+\delta'_{3})\ ,     
\end{eqnarray}
where $M_{1,2,3}$ are the bino, wino and gluino masses, respectively, and
we assume that $ -1 \leq \delta'_{i} \leq 1 $.

Let us discuss now which values of the parameters are interesting
in order to increase the cross section, with respect to the
universal case $\delta'_i=0$. In this sense, it is worth noticing that
$M_3$ appears in the RGEs of squark masses, so e.g.
their contribution proportional
to the top Yukawa coupling in the RGE of $m_{H_u}^2$ will
do this less negative if $M_3$ is small. Therefore $\mu^2$ becomes smaller.

\begin{figure}[t]
\begin{center}
\epsfig{figure=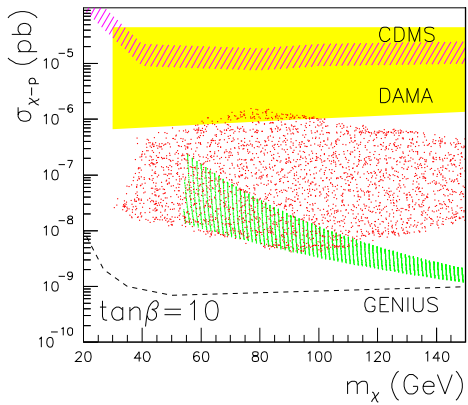,
height=7cm}
\end{center}

\vspace{-0.5cm}

\caption{
The same as in Fig.~\ref{ellisfig}b
but only for $\tan{\beta}=10$ and using non-universal soft 
gaugino masses 
as explained in the text. The universal case (green region) 
is shown for comparison.
\label{nunivg.eps}}
\end{figure}

Taking into account this effect, 
we show in Fig.~\ref{nunivg.eps},
for $\tan\beta=10$,
a scatter plot of
$\sigma_{\tilde{\chi}_{1}^{0}-p}$ as a function of $m_{\tilde{\chi}_{1}^{0}}$
for the following scanning of the parameters:
$ -1 \leq \delta'_{1,2} \leq 1 $
and $ -2/3 \leq \delta'_{3} \leq 0 $ (where the lower bound
is due to the experimental bound on gluino masses), imposing 
$M_3<M_1, M_2$.
We
fix 
$M_{1/2}= 225$ GeV and $A=M_{1}$.
As in the previous figures
we are 
taking 30 GeV $\leq m_0\leq 550$ GeV.
We also superimpose for comparison the region (green area) 
obtained in Fig.~\ref{ellisfig2} for $\tan\beta=10$
with universality, $\delta'_i=0$.
Clearly non-universal gaugino masses produce a large 
increment in the cross section.

Let us finally remark that the upper and right hand side regions with
large cross sections in the
figure correspond to points of the parameter space fulfilling 
$M_2<M_1$ (and $M_3$ close to its lower bound).
As discussed in Section~2,
for
low or moderate $\tan\beta$ the lightest neutralino is mainly gaugino $P>0.9$,
and in particular bino since $N_{11}\gg N_{12}$. However, relaxing the
universality condition for gaugino masses, $M_2<M_1$
will produce $N_{11}<N_{12}$, and therefore
the lightest neutralino will have an important wino 
component\footnote{It is worth noticing that this situation
also arises in anomaly-mediated SUSY-breaking scenarios, where
the cross section can also be increased \cite{anomaly,anomaly2}.}.
Since the latter couples through $SU(2)$
interactions, unlike the bino component which couples through $U(1)$
ones, the cross section increases.\\

\hspace{-0.65cm}{\it $(iii)$ Non-universal gaugino and scalar masses}\\

Given the above situation concerning the enhancement of
neutralino-proton cross sections through non-universality 
for moderate $\tan\beta$, it is worth analyzing in
principle the combination of both possibilities,
non-universal gaugino and scalar masses.
However 
the consequence
for the cross section of combining the parameters studied in
{\it (i)} and {\it (ii)} is not very different from the one obtained
in Fig.~\ref{nunivg.eps}.
We have explicitly checked that the qualitative pattern is basically
the same. 

\subsection{\normalsize Scenario with intermediate unification scale}

In the above subsections the analyses were performed assuming the
unification scale
$M_{GUT} \approx 10^{16}$ GeV, as is usually done in the SUSY literature.
However, it was recently realized that 
the string scale may be anywhere between the weak and the Plank 
scale \cite{Lykken}-\cite{weakandstronghete}.
To use the value of the initial scale,
say $M_I$, as a free parameter for the running of the soft terms
is particularly interesting since 
there are several arguments in favour of SUSY scenarios
with scales 
$M_I\approx 10^{10-14}$ GeV.

First, these scales were suggested \cite{stronghete,typeIinter}
to explain many experimental observations as
neutrino masses or the scale for axion physics. 
Second, with the string scale of the order of
$10^{10-12}$ GeV
one is able
to attack the hierarchy problem of unified theories 
without invoking any hierarchically suppressed non-perturbative
effect \cite{stronghete,typeIinter}. Third,
for intermediate scale scenarios
charge and color breaking constraints,
which are very strong with the usual scale $M_{GUT}$ \cite{servidor},
become less important \cite{Allanach}.
There are other arguments in favour 
of scenarios with initial scales $M_I$ smaller
than $M_{GUT}$.
For example these scales
might also
explain the observed ultra-high energy ($\approx 10^{20}$ eV) cosmic rays
as products of long-lived massive string mode 
decays \cite{stronghete,typeIinter} (see ref.\cite{cosmic} for more
details about this possibility). Besides,
several models of chaotic inflation favour also these scales \cite{caos}.

Inspired by these scenarios, 
it was pointed out 
\cite{muas}-\cite{nosotros} 
that the
neutralino-proton cross section $\sigma_{\tilde{\chi}_1^0-p}$
is very sensitive to the variation of the initial scale 
for the running of the soft terms.
In particular, intermediate unification scales were considered.
For instance, by taking $M_I=10^{10-12}$ GeV rather than 
$M_{GUT}$, regions in the parameter space of mSUGRA have been 
found 
where  $\sigma_{\tilde{\chi}_1^0-p}$
is in the expected range of sensitivity of present detectors,
and this
even for moderate values of $\tan\beta$ ($\tan\beta\gsim 3$). 
This analysis 
was performed \cite{muas} in the universal scenario for the soft terms. 
In contrast, in the usual case with initial scale at $M_{GUT}$,
this large cross section is achieved only for $\tan\beta \gsim 20$,
as discussed in Subsection~3.1.

\begin{figure}[t]
\begin{center}
\epsfig{file= 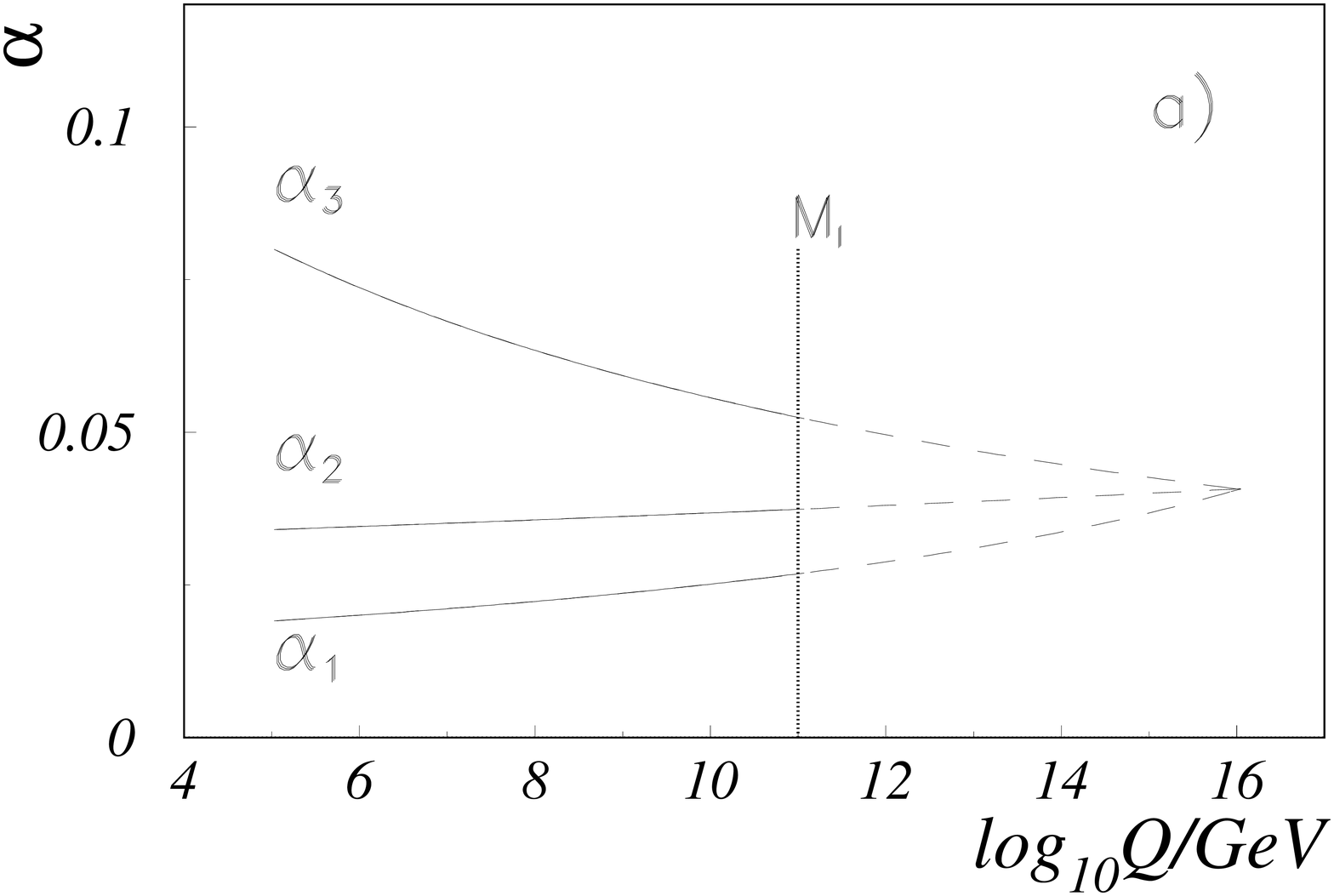,width=5.75cm,height=4.5cm,angle=0}
\hspace{0.5cm}\epsfig{file= 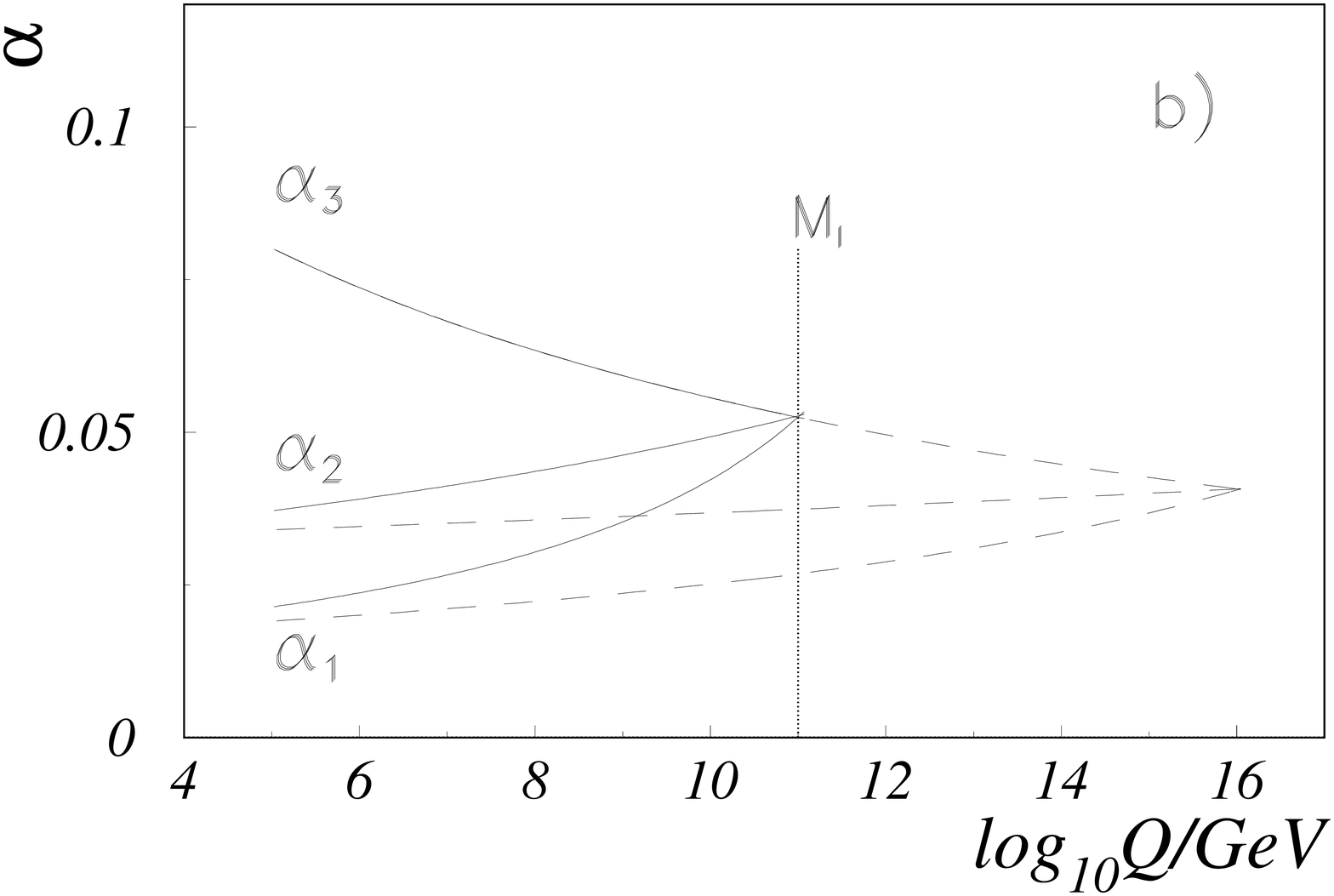, width=5.75cm, height=4.5cm,angle=0}
\end{center}

\vspace{-0.5cm} 

\caption{Running of the gauge couplings with energy, shown
with solid lines, assuming
(a) non universality and (b) universality of couplings at the
initial scale $M_I$. 
For comparison the usual running of the MSSM couplings
is also shown with dashed lines.}
\label{uni}
\end{figure}

Before trying to understand this result, let us discuss what we
mean by an intermediate unification scale. 
Concerning this point
two possible scenarios
are schematically shown in Fig.~\ref{uni} for the example
$M_I= 10^{11}$ GeV. In scenario (a) 
the gauge couplings are non universal,
$\alpha_i\neq\alpha$, and their values
depend on the initial scale $M_I$ chosen.
An interesting proposal in order to obtain this scenario
in the context of
type I string models is the following. 
If the standard model comes from the
same collection of D-branes, 
stringy corrections might change the boundary conditions at the string scale
$M_I$ to mimic the effect of field theoretical logarithmic 
running \cite{Rigolin,mirage}.
Another possibility giving rise to a similar result
may arise when
the gauge groups came from different types of D-branes.
Since different D-branes have associated different couplings,
this implies the non universality of the gauge couplings. 
We will discuss this possibility in some detail below.

On the other hand, scenario (b) with gauge coupling 
unification at $M_I$, $\alpha_i=\alpha$, can
be obtained with 
the addition of extra fields
in the massless spectrum. For the example of the figure these are
doublets and singlets under the standard model gauge group.

As we will see below, the values of the gauge coupling constants at the
intermediate
scale will be important in the computation of the cross section, and
scenario (a) will be more interesting than (b).

\begin{figure}[t]
\begin{center}
\epsfig{file= 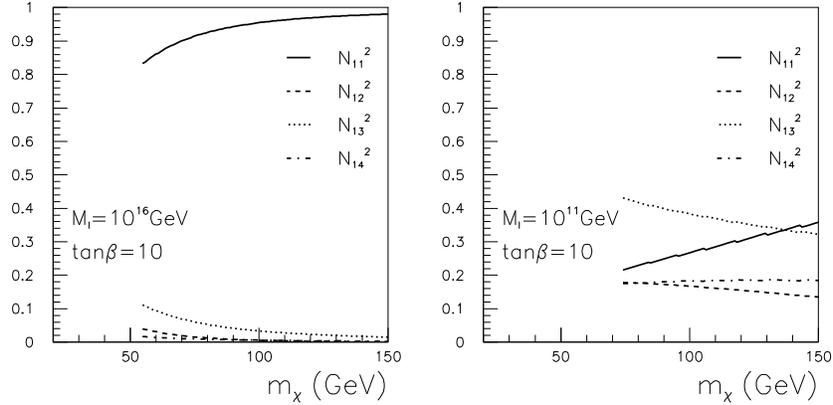, 
height=7cm,angle=0}
\end{center}

\vspace{-0.5cm}

\caption{Gaugino-Higgsino components-squared
of the lightest neutralino as a function of its mass for
the unification scale, $M_I=10^{16}$ GeV, and for the intermediate scale,
$M_I=10^{11}$ GeV.
}
\label{N_1i}
\end{figure}

Let us now come back to the issue of the variation of the
cross section with the initial scale.
The fact that smaller initial scales imply a larger neutralino-proton
cross section can be understood from the variation in the value of 
$\mu$ with $M_I$.
One observes that, for $\tan\beta$ fixed, the smaller the initial
scale for the running is, the smaller the numerator in the
first piece of eq.(\ref{electroweak}) becomes. 
This can be understood qualitatively from Fig.~\ref{run}. 
Clearly, the smaller the initial scale is, the shorter the
running becomes. As a consequence, 
also the less important the positive(negative) contribution 
$m_{H_d}^2$($m_{H_u}^2$) to $\mu$ in eq.(\ref{electroweak}) becomes.
Thus 
$|\mu|$ decreases. 

\begin{figure}[t]
\begin{center}
\epsfig{file= 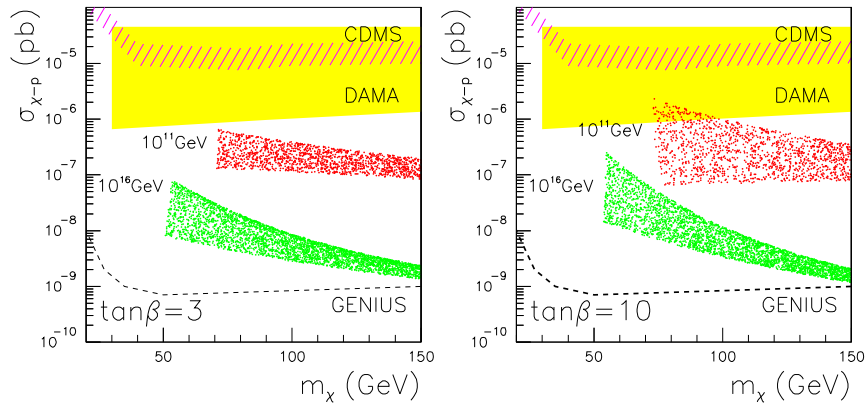, 
height=7cm}
\end{center}

\vspace{-0.5cm}

\caption{The same as in Fig.~\ref{ellisfig}b
but for two values of the initial scale,
$M_I=10^{16}$ GeV and $10^{11}$ GeV, and for $\tan\beta=3$ and 10. 
}
\label{sigmaM_I}
\end{figure}

As discussed at the beginning of this Section~3, now
the Higgsino
components
of the lightest neutralino, $N_{13}$ and $N_{14}$ in eq.(\ref{lneu}),
increase 
and therefore  
the spin independent cross section also 
increases \cite{muas}.
This is shown  in Figs.~\ref{N_1i} and \ref{sigmaM_I}.
In fact these figures correspond to the scenario 
in Fig.~\ref{uni}a with non-universal gauge couplings.
For the scenario in Fig.~\ref{uni}b, 
only with $\tan\beta\gsim 20$ one obtains
regions consistent with DAMA limits.
One of the reasons being that now
$\alpha_2 (M_I)$ and $\alpha_1 (M_I)$ are bigger than in scenario (a) 
and therefore the low-energy bino and wino masses appearing
in eq.(\ref{mass}) are smaller.
As a consequence 
the increment of the cross section is less important.

It is also worth noticing
that, for any fixed value of $M_I$, the larger
$\tan\beta$ is, the larger
the Higgsino contributions become.
The discussion concerning this point in Subsection~3.1 for $M_{GUT}$ is also
valid for any initial scale $M_I$.

In Fig.~\ref{N_1i}, for $\tan\beta=10$ and $m_0=150$ GeV, 
we exhibit the gaugino-Higgsino components-squared $N_{1i}^2$ 
of the lightest neutralino as a function of its mass $m_{\tilde{\chi}_1^0}$
for 
two different values of 
the initial scale, 
$M_I=10^{16}$ GeV  $\approx M_{GUT}$ and 
$M_I=10^{11}$ GeV. 
Clearly, the smaller the scale is, the larger 
the Higgsino components become. 
%
%
For
$M_I=10^{11}$ GeV, e.g. the Higgsino contribution $N_{13}$  
becomes important 
and even dominant for 
$m_{\tilde{\chi}_1^0}\lsim 140$ GeV.

%
%


The consequence of the above results on the cross section is shown in 
Fig.~\ref{sigmaM_I}, where
the cross section as a function
of the lightest neutralino mass $m_{\tilde{\chi}_1^0}$ is plotted.
In particular we are comparing the result for the scale $M_I=M_{GUT}$
studied in Figs.~\ref{ellisfig} and \ref{ellisfig2} with the result for 
the intermediate scale $M_I=10^{11}$ GeV.
For instance, when $m_{\tilde{\chi}_1^0}=100$ GeV, 
$\sigma_{\tilde\chi_1^0-p}$ 
for $M_I=10^{11}$ GeV is two orders of magnitude larger
than for $M_{GUT}$.
In particular, for $\tan\beta=3$, one finds 
$\sigma_{\tilde\chi_1^0-p} \lsim 10^{-7}$ pb 
if the initial scale is $M_I=10^{16}$ GeV.
However $\sigma_{\tilde\chi_1^0-p} \lsim 10^{-6}$ GeV is possible if 
$M_I$ decreases. 

As mentioned before, the larger
$\tan\beta$ is, the larger the Higgsino contributions become, and
therefore the cross section increases.
For $\tan\beta=10$ we see in Fig.~\ref{sigmaM_I}
that the range 70 GeV $\lsim m_{\tilde{\chi}_1^0}\lsim$ 100 GeV is now
consistent with DAMA limits. 

Let us remark that these figures have been obtained \cite{muas}
taking $30\lsim m_0\lsim 550$ GeV and $A_0=M_{1/2}$ as in previous sections.
In any case, as mentioned also in Section~2, the cross section is 
not very sensitive to the specific values
of $A_0$. In particular it was checked that 
this is so for 
$\mid A_0/M_{1/2}\mid \lsim 1$.
For example, relation $A_0=-M_{1/2}$ 
is particularly interesting since it arises naturally
in several string models \cite{dilaton,Rigolin}. 


Let us finally recall that
the above computations have been carried out for the case of
universal soft terms. This is not only the most simple
possibility in the framework of SUGRA, but  
is also allowed in the context of superstring models. This is e.g.
the case of the dilaton-dominated SUSY-breaking 
scenario \cite{dilaton}
or weakly and strongly coupled heterotic models with one K\"ahler 
modulus \cite{softM}.
In this sense the analysis of neutralino-nucleon
cross sections of those models 
is included in the above analyses. This is exactly true for
$M_I=M_{GUT}$. For intermediate unification scale,
note that we are assuming gaugino mass universality
at the high energy scale, 
although in this scenario gauge couplings do not unify. 
This situation is in principle
possible in generic supersymmetric models, however it is not so natural in
supersymmetric models from supergravity 
where gaugino masses and gauge couplings
are related through the gauge kinetic function. 
Since an explicit string construction with nonuniversal
gauge couplings and gaugino masses will be analyzed in detail below,
we have chosen to simplify the discussion here
assuming gaugino mass universality.

Obviously, 
following the discussion of Subsection~3.3,
non universality of the soft terms in 
addition to intermediate scales
may introduce more flexibility in the computation.
In particular, decreasing $|\mu|$
in order to obtain regions in the parameter space giving rise to  
cross sections compatible with the sensitivity
of current detectors, may be easier.\\

\hspace{-0.65cm}{\it  D-brane scenarios}\\

D-brane constructions are explicit scenarios where both situations 
mentioned above,
non-universality and intermediate scales, may occur.
The first attempts to study dark matter within these constructions
were carried out in scenarios with the unification scale  
$M_{GUT} \approx 10^{16}$ GeV
as the initial scale \cite{khalil,Nath2,Arnowitt2}
and dilaton-dominated SUSY-breaking scenarios 
with an intermediate scale as the initial scale \cite{bailin}.
However,
the important issue of the D-brane origin of the $U(1)_Y$ gauge group
as a combination of other $U(1)$'s 
and its influence on the matter distribution in these scenarios
was not included in the above analyses.
When this is taken into account, interesting results are 
obtained \cite{nosotros}. In particular, 
scenarios with the gauge group and particle content of the
SUSY standard model lead naturally to intermediate values for the
string
scale, in order to reproduce the value of gauge couplings
deduced from experiments. In addition, the soft terms 
turn out to be generically non universal.
Due to these results, 
large cross sections
in the small $\tan\beta$ regime
can be obtained.

Let us consider for example a type I string scenario \cite{nosotros}
where the gauge group 
$U(3)\times U(2)\times U(1)$, giving rise to
$SU(3)\times SU(2)\times U(1)^3$, arises
from three different types of D-branes, and therefore
the gauge couplings are
non-universal
as in Fig.~\ref{uni}a\footnote{To be precise, the running of the 
$U(1)_Y$ gauge coupling is not exactly like in the figure since, due to the
D-brane origin of the $U(1)$ gauge groups, relation (\ref{couplings})
must be fulfilled. See in this respect Fig.~2 in ref.\cite{nosotros}.}.
Other examples with the standard model gauge group embedded in D-branes in 
a different way can be found in ref.\cite{nosotros}.
Here
$U(1)_Y$ is a linear combination
of the three $U(1)$ gauge groups arising from $U(3)$, $U(2)$ and
$U(1)$ within the three different D-branes. This implies
\begin{equation}
\frac{1}{\alpha_Y(M_I)} =     
\frac{2}{\alpha_1(M_I)} + \frac{1}{\alpha_2(M_I)} 
+ \frac{2}{3 \alpha_3(M_I)}\ ,
\label{couplings}
\end{equation}
where $\alpha_k$ correspond to the gauge couplings of the $U(k)$ branes.
As shown in ref.\cite{nosotros},  $\alpha_1(M_I) = 0.1$ leads
to the string scale $M_I = 10^{12}$ GeV. 
On the other hand,
the extra $U(1)$'s are anomalous and therefore the associated gauge
bosons
have masses of the order of $M_I$.

The analysis of the soft terms has been done under
the assumption that only the
dilaton ($S$) and moduli ($T_i$) fields contribute to SUSY
breaking and it has been found  that these soft terms  are generically
non-universal.
Using the standard parameterization \cite{dilaton} 
\begin{eqnarray}
&& F^S= \sqrt{3} (S+S^*) m_{3/2} \sin \theta\;, \nonumber \\
&& F^i= \sqrt{3} (T_i+T^*_i) m_{3/2} \cos \theta\; \Theta_i\;,
\end{eqnarray}
where
$i=1,2,3$ labels the three complex compact dimensions, and
the angle $\theta$ and the $\Theta_i$ with $\sum_{i} |\Theta_i|^2=1$,
just parameterize the direction of the goldstino in the $S$, $T_i$ field
space, one is able to obtain the following soft terms \cite{nosotros}.
The gaugino masses associated to the three gauge groups of the
standard model are given by
%
\bea
M_3 & = & \sqrt{3} m_{3/2} \sin \theta \ , \nn\\
M_{2} & = & \sqrt{3}  m_{3/2}\ \Theta_1 \cos \theta  \ , \nn\\
M_{Y} & = &  \sqrt{3}  m_{3/2}\ \alpha_Y (M_I)
\left(\frac{2\ \Theta_3 \cos \theta}{\alpha_1 (M_I)}
+\frac{\Theta_1 \cos \theta}{\alpha_2 (M_I)}
+\frac{2\ \sin \theta}{3 \alpha_3 (M_I)}
\right)\ .
\label{gaugino1}
\eea
The soft scalar masses of the three families are given by
\begin{eqnarray}
m^2_{Q_L} & = & m_{3/2}^2\left[1 -
\frac{3}{2}  \left(1 - \Theta_{1}^2 \right)
\cos^2 \theta \right] \ , \nn \\
m^2_{d_R} & = & m_{3/2}^2\left[1 -
\frac{3}{2}  \left(1 - \Theta_{2}^2 \right)   
\cos^2 \theta \right] \ , \nn \\
m^2_{u_R} & = & m_{3/2}^2\left[1 -
\frac{3}{2}  \left(1 - \Theta_{3}^2 \right)
\cos^2 \theta \right] \ , \nn \\
m^2_{e_R} & = & m_{3/2}^2\left[1- \frac{3}{2}
\left(\sin^2\theta + \Theta_{1}^2 \cos^2\theta  \right)\right] \ , \nn \\
m^2_{L_L} & = & m_{3/2}^2\left[1- \frac{3}{2}
\left(\sin^2\theta + \Theta_{3}^2 \cos^2\theta  \right)\right] \ , \nn \\
m^2_{H_u} & = & m_{3/2}^2\left[1- \frac{3}{2}
\left(\sin^2\theta + \Theta_{2}^2 \cos^2\theta  \right)\right] \ , \nn \\
m^2_{H_d} & = & m^2_{L_L} \;,     
\label{scalars1}
\end{eqnarray}
where e.g. $u_R$ denotes the three family squarks $\tilde{u}_R$, 
$\tilde{c}_R$, $\tilde{t}_R$.
Finally the trilinear parameters of the three families are
\begin{eqnarray}
A_{u} & = &  \frac{\sqrt 3}{2}m_{3/2}
   \left[\left(\Theta_{2} - \Theta_1
 - \Theta _{3}  \right) \cos\theta
- \sin\theta \; \right] \ ,
\nn \\
A_{d} & = &  \frac{\sqrt 3}{2}m_{3/2}
   \left[\left(\Theta_{3} - \Theta_1
  - \Theta _{2}  \right) \cos\theta
- \sin\theta \; \right] \ ,
\nn \\
A_{e} & = &  0\; .
\label{trilin11}
\end{eqnarray}

Although these formulas for
the soft terms imply that one has in principle
five free parameters, $m_{3/2}$, $\theta$ and  $\Theta_i$ with $i=1,2,3$,
due to relation $\sum_i |\Theta_i|^2=1$ only four of them are
independent.
In the analysis the parameters $\theta$ and $\Theta_i$
are varied in the whole allowed range, $0\leq \theta\leq 2\pi$,
$-1\leq\Theta_i\leq 1$. For 
the gravitino mass, 
$m_{3/2}\leq 300$ GeV is taken.
Concerning Yukawa couplings, their values are fixed imposing 
the correct fermion mass spectrum at low energies, i.e.,
one is assuming that Yukawa structures of D-brane scenarios
give rise to those values.

\begin{figure}[t]
\begin{center}
\epsfig{file= 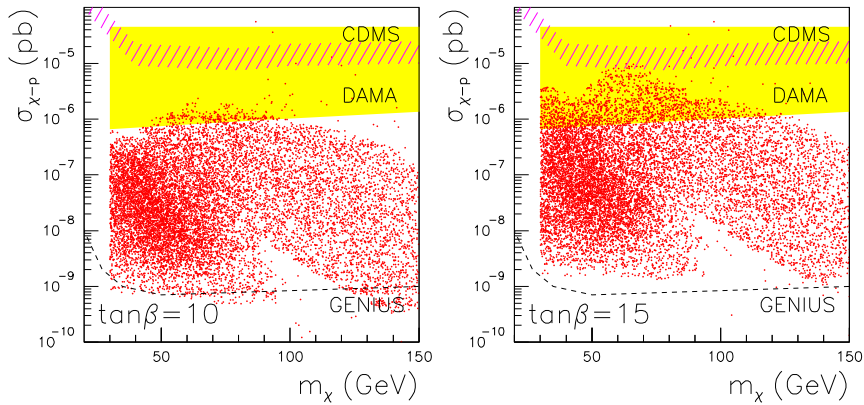, 
height=7cm
}
\end{center}
\vspace{-1cm}
\caption{The same as in Fig.~\ref{ellisfig}b but for 
the D-brane scenario with 
the string scale $M_I=10^{12}$ GeV
discussed
in the text, and for $\tan\beta= 10$ and 15.}
\label{dbrane}
\end{figure}

Fig.~\ref{dbrane} displays a scatter plot of 
$\sigma_{\tilde\chi_1^0-p}$ as a function of the
neutralino mass $m_{\tilde\chi_1^0}$ for a scanning of the parameter space
discussed above. 
Two different values of $\tan\beta$, 10 and 15, are shown.
LEP and Tevatron bounds
on SUSY masses are included as in the previous sections. 
They forbid e.g. values of
$m_{3/2}$ smaller than 170 GeV.
Although bounds coming from CLEO $b\rightarrow s\gamma$ branching
ratio measurements are not included in the figures, one can check
explicitly that their qualitative patterns are not modified.
It is worth noticing that
for $\tan\beta =10$ there are regions of the parameter
space consistent with DAMA limits.
In fact, one can check that $\tan\beta > 5$ is enough to
obtain compatibility with DAMA.
Since the larger $\tan\beta$ is, the
larger the cross section becomes, for $\tan\beta =15$ these regions
increase.

Let us recall that both plots in the figure are obtained taking
$m_{3/2}\leq 300$ GeV, which corresponds to squark masses 
smaller than $500$ GeV  at low energies.
Larger values of $m_{3/2}$ will always produce cross sections 
below DAMA limits. In particular, 
the right hand side and bottom of the plots will also be filled
with points.
Cross sections below
projected GENIUS limits will be possible for both figures.
On the other hand, it is worth mentioning that the isolated 
points in the plots with, in general, very large values of the cross section
correspond to values of the lightest stop mass extremely close to the
mass of the LSP, 
in particular $(m_{\tilde t}-m_{\tilde\chi_1^0})/m_{\tilde t}<0.01$.

Finally, let us mention that scenarios where all gauge groups of the 
standard model are
embedded
within the same set of D-branes, and therefore with gauge coupling
unification, are also possible.
However, unlike the previous scenario, 
now $\tan\beta > 20$ is necessary in order to obtain regions
consistent with DAMA limits \cite{nosotros}.
As discussed above in the context of mSUGRA 
this is due to the different values of the $\alpha$'s at the string scale  
in both types of scenarios. 

\section{\large Relic neutralino density versus cross section}

As discussed in the Introduction, current dark matter detectors
are sensitive to a neutralino-proton cross section around $10^{-6}$ pb.
This value is obtained taking into account, basically, that the
density of dark matter in our Galaxy, which follows from the 
observed rotation curves, is 
$\rho_{DM}\approx 0.3$ GeV/cm$^3$.
Thus in this work we were mainly interested in reviewing the
possibility of obtaining such large cross sections in the
context of SUSY scenarios. In order to compute the cross section
only simple field theory techniques are needed,
no cosmological assumptions about the
early Universe need to be used.

On the other hand, such
cosmological assumptions indeed must be taken into account 
when computing the amount of relic neutralino density
arising from the above scenarios.
Generically, one obtains \cite{kami}
%
\begin{equation}
\Omega_{\tilde{\chi}_1^0} h^2 \simeq \frac{C}
{<\sigma_{\tilde{\chi}_1^0}^{ann}. v>}
\ ,
\label{assumption}
\end{equation}
where $\sigma_{\tilde{\chi}_1^0}^{ann}$ is the  
cross section for annihilation of a pair of neutralinos
into standard model particles, $v$ is the 
relative velocity between the two neutralinos, 
and $<..>$ denotes thermal averaging.
The constant $C$ involves factors of Newton's constant, the temperature of the 
cosmic background radiation, etc. 
Then one may compare this result with dark matter observations
in the Universe.
Let us then discuss briefly the effect of relic neutralino 
density bounds on cross sections.

The most robust evidence for the existence of dark matter comes from 
relatively small scales.
Lower limits inferred
from the flat rotation curves of spiral Galaxies \cite{kami,salucci} are 
$\Omega_{DM}\ h^2\gsim 0.01-0.05$, where
$h$ is the reduced Hubble constant. 
On the opposite side, observations at large scales,
$(6-20)\ h^{-1} $ Mpc, have provided 
estimates \cite{freedman} 
of $\Omega_{DM} h^2\approx 0.1-0.6$, but values as low as 
$\Omega_{DM} h^2\approx 0.02$ have also been quoted \cite{kaiser}.
Taking up-to-date limits on $h$, the 
baryon density from nucleosynthesis and overall matter-balance  
analysis one is able to obtain a favoured 
range \cite{Sadoulet,primack00,NEW},
$0.01\lsim \Omega_{DM} h^2 \lsim 0.3$ (at $\sim 2\sigma$ CL).
Note that  conservative 
lower limits in the small and large scales are 
of the same order of magnitude.

As is well known, for $\sigma_{\tilde{\chi}_1^0}^{ann}$ 
of the order 
of a weak-process cross section, $\Omega_{\tilde{\chi}_1^0}$ obtained
from eq.(\ref{assumption}) is 
within the favoured range discussed
above \cite{kami}. 
This is precisely the generic case when the lightest neutralino
is mainly bino. Then, the neutralino-nucleus cross section is
of the order of 1 pb,
i.e. $\sigma_{\tilde{\chi}_1^0-p}\approx 10^{-8}$ pb, 
and therefore it is natural to obtain
that neutralinos annihilate with very roughly
the weak interaction strength. These cross sections were
discussed in Section 2 where
low and moderate values of $\tan\beta$ within mSUGRA were considered.
In fact, for these cross sections, there is always
a set of parameters which yield $0.1<\Omega_{\tilde\chi_1^0}h^2<0.3$.
This analysis, including a complete treatment of coannihilations
was carried out in refs.\cite{Ellis,Ellisco}.

On the other hand, in this review we were interested in larger
neutralino-nucleon cross sections in order to be in the range
of sensitivity of current dark matter detectors.
It is then expected that such high neutralino-proton cross sections
$\sigma_{\tilde\chi_1^0-p}\approx 10^{-6}$ pb,
as those presented in Section~3, will correspond to relatively low
relic neutralino densities.
This is in general the situation. There is always a set
of parameters for which the value of the relic density
is still inside the ranges we considered above when discussing the
observational bounds, but it is generically close to the
conservative lower bound.
The analysis of the relic neutralino density for
scenarios with large $\tan\beta$ and non-universal scalar masses 
was carried out in refs.\cite{Bottino,Arnowitt,Mario,Arnowittco}.
Scenarios with non-universal gaugino masses were studied in
refs.\cite{Yamaguchi,Nelson}.
An analysis of the relic density for focus point supersymmetry 
can be found in ref.\cite{focus}. Discussions for
scenarios with intermediate unification scales can be found in
refs.\cite{muas}-\cite{nosotros}\footnote{Let us remark, however,
that scenarios with intermediate scales might give rise to cosmological
results different from the usual ones 
summarized in eq.(\ref{assumption}) \cite{preparation}.}. 

Finally, in case of 
preferring the stronger lower bound
$\Omega_{DM}h^2>0.1$, let us mention
the possibility that not all the dark matter
in our Galaxy are neutralinos. Then 
$\Omega_{\tilde\chi_1^0}<\Omega_{DM}$,
and therefore $\Omega_{\tilde\chi_1^0}<0.1$ is possible. However,
due to the corresponding reduction in the density of neutralinos in the 
Galactic halo, the neutralino-proton
cross section should be increased in order to maintain the 
experimental detection rates \cite{Ellis,elotrodeellis,Nojiri}.

\section{\large Final comments and outlook}

There is overwhelming evidence that most of the matter in the Universe
is dark matter.
In the present paper we have reviewed the direct detection of 
supersymmetric dark matter in the light of recent experimental efforts.
In particular, DAMA collaboration using a NaI detector
has reported recently \cite{experimento1}
data favouring the existence of a WIMP signal in their search for
annual modulation. They require a large cross section of the order of
$10^{-6}$ pb. 
We have observed that there are regions in the parameter space
of SUGRA scenarios \cite{Bottino}-\cite{nosotros} 
where such a value can be obtained, although it is fair to say that
smaller values
can also be obtained and even more easily.
The latter result may be important since 
CDMS collaboration using a germanium detector
has reported a null result for part of the region explored by DAMA. 
Clearly, more sensitive detectors
producing further data are needed to solve this contradiction.
Fortunately, many dark matter detectors are being projected.
This is the case e.g. of DAMA 250 kg. and CDMS Soudan, but
particularly interesting is the projected GENIUS \cite{GENIUS}
detector where values of the cross section as low as 
$10^{-9}$ pb will be accesible. 

In summary, underground physics as the one discussed here in order
to detect dark matter is crucial. Even if neutralinos are discovered
at future particle accelerators such as LHC, only their direct detection
due to their presence in our galactic halo will confirm that they
are the sought-after dark matter of the Universe.
\\

\noindent {\bf Acknowledgments}

\noindent
We thank E. Gabrielli and E. Torrente-Lujan
as co-authors of some of the works reported in
this review. D.G. Cerde\~no acknowledges the financial support
of the Comunidad de Madrid through a FPI grant.
The work of S. Khalil was supported by PPARC.
The work of C. Mu\~noz was supported in part by the Ministerio de 
Ciencia y Tecnolog\'{\i}a, and the European Union under contract 
HPRN-CT-2000-00148.

\end{document}